\documentclass[letterpaper, 10 pt, conference]{ieeeconf}  % Comment this line out if you need a4paper
\usepackage{amsmath,amssymb,amsfonts}
\usepackage{graphicx}
\usepackage{psfrag}

\IEEEoverridecommandlockouts                              % This command is only needed if 
\title{\LARGE \bf
Modeling Physical Activity Impact on Glucose Dynamics in People with Type 1 Diabetes for a Fully Automated Artificial Pancreas}

\author{Mehrad Jaloli and Marzia Cescon$^{*}$  %<-this % stops a space
\thanks{The authors are with the Dept. of Mechanical Engineering, University of Houston, Houston TX. $^{*}$ Corresponding author: {\tt\small mcescon2@uh.edu}}%
\thanks{This work was supported by the University of Houston through a startup grant.}
}

\begin{document}

\maketitle
\thispagestyle{empty}
\pagestyle{empty}

%%%%%%%%%%%%%%%%%%%%%%%%%%%%%%%%%%%%%%%%%%%%%%%%%%%%%%%%%%%%%%%%%%%%%%%%%%%%%%%%
\begin{abstract}
In this paper, models of the blood glucose (BG) dynamics in people with Type 1 diabetes (T1D) in response to moderate intensity aerobic activity are derived from physiology-based first principles and system identification experiments. We show that by enhancing insulin-dependent glucose utilization by the tissues in two phases, a rapid short-term increase in insulin-independent glucose clearance and augmented glucose uptake, and a long-term sustained increase sensitivity to insulin action, a metabolic model able to reproduce the effects of activity on glucose disposal is obtained. Second, a control-oriented transfer function model is proposed to predict the BG response to an exercise bout modeled as a step change in heart rate (HR). Results comparing model predictions with actual patients data collected in a series of experimental sessions including physical activity (PA) are presented. The findings will contribute to the design of a fully automated closed-loop for improved glucose control in conditions of daily life for people with T1D.
\end{abstract}

%%%%%%%%%%%%%%%%%%%%%%%%%%%%%%%%%%%%%%%%%%%%%%%%%%%%%%%%%%%%%%%%%%%%%%%%%%%%%%%%
\section{INTRODUCTION}
%Here comes the introduction of this paper dealing with modeling of the effect of moderate intensity aerobic activity on blood glucose dynamics in people with T1D, and implications for control design. 
Type 1 diabetes (T1D) is a chronic disorder of impaired glucose metabolism characterized by the lack of insulin secretion by the pancreas. Exogenous insulin replacement is the hallmark of the therapy for the people affected, to mimic as closely as possible the physiological insulin secretion pattern of the healthy individual, and thus reducing markedly the long-term complications of the chronic hyperglycemia caused by insufficient insulin~\cite{2018PAtG}. The American Diabetes Association recommends regular physical exercise of at least 150 minutes per week for the numerous health and wellbeing effects associated with it~\cite{2018ADA}. However, adapting insulin delivery both during and after physical exercise is particularly difficult because in people with T1D, glycemic response to exercise is highly variable, both within and among individuals, often increasing the risk for hypoglycemia and hyperglycemia, the reason(s) behind this variability being largely unknown. As a result, for most people, diabetes management is perceived as a barrier to exercise~\cite{brazeau2008}.
While continuous glucose monitors (CGMs) and automated insulin delivery (AID) systems including the artificial pancreas (AP) hold promise for improved glycemic control during and after exercise, the lack of accurate models able to describe the effect of physical activity (PA) on glucose metabolism is a key limiting factor to the development of insulin treatment strategies applicable to conditions of daily life. Against this background, the objective of this study is twofold. Firstly, we propose an extension to a well known simulation model of the glucoregulatory system able to accurately reproduce the effect of moderate intensity aerobic activity on glucose dynamics compared to actual data; secondly, we present a simple first-order transfer function with time delay modeling the PA disturbance useful for control design. 

A few models describing the effect of exercise on glucose excursions are reported in the literature. Roy and Parker~\cite{roy2007dynamic} developed a PA model as an extension to the Bergman Minimal Model~\cite{bergman1979identification}, relating the maximum rate of oxygen consumption (VO$_2$max) to glucose uptake in the periphery; in~\cite{breton2008} Breton proposed a parsimonious exercise model linking the change in insulin action and glucose effectiveness to heart rate (HR); later Dalla Man et. al.~\cite{dallaMan2009} developed and tested three extension of~\cite{breton2008} and included them into a simulation model of a glucose–insulin system~\cite{dallaMan2007}. More recently, Alkhateeb et. al.~\cite{alkhateeb2021modelling} developed six variations of the Bergman Minimal Model~\cite{bergman1979identification} to describe the impact of moderate exercise on T1D glucose dynamics fitting their models to data collected in closed-loop during continuous and interval exercise. In all previous studies, the authors presented \textit{in-silico} experiments only without comparing model predictions back-to-back with actual data. When trying to replicate clinical protocols, assessment were only qualitative and limited to the exercise session alone, disregarding post-exercise effects. As far as data-driven black-box models are concerned, some recent contributions include using data collected with a wearable device as a proxy for exercise to identify linear model for prediction of BG dynamics in activity of daily life~\cite{faccioli2018black}; Xie et. el.~\cite{xie2019data} proposing a nonlinear ARMAX model where the effect of PA on the BG variation was characterized through a linear regression with respect to the activity input, and a bilinear regression with respect to the activity and insulin inputs; and~\cite{isuru} in which Dasanayake and colleagues estimated empirical dynamic models of the BG dynamics in reponse to PA measured by commercially available activity monitors, exploiting subspace-based methods for system identification. %the authors either presented simulation results only, or when comparing their simulated data with the actual data from clinical trial, the comparison was limited to the exercise session alone, which was used to fit the model, without considering the long-lasting effects of exercise on insulin sensitivity and glucose uptake, ultimately perturbing glucose dynamics for up to 16 hours.  

The remainder of the paper is
organized as follows. In the next section, the experimental conditions and clinical data acquisition in a series of experimental trials are presented; models of the effect of PA on BG dynamics in people with T1D are introduced in Sec. \ref{sec:modeling}. Sec. \ref{sec:results} is devoted to the presentation of the results while Sec.~\ref{sec:conclusions} summarizes the findings and concludes the paper.
%Sec. \ref{sec:control_design} deals
%with the design of a decision support system for automatic titration and insulin dosing during exercise;
%%%%%%%%%%%%%%%%%%%%%%%%%%%%%%%%%%%%%%%%%%%%%%%%%%%%%%%%%%%%%%%%%%%%%%%%%%%%%%%%
\section{Experimental Conditions and Clinical Data Acquisition}\label{experimental_cond}
Data acquisition was performed in a series of clinical visits within a randomized, 4-way crossover trial conducted in compliance with the ethical principles in the Declaration of Helsinki and with the standards of Good Clinical Practice. The protocol
received Institutional Review Board approval at the clinical site, and was registered at clinicaltrials.gov (NCT02660242). Eligible subjects were between 18 and 65 years of age with T1D for at least two years, using an insulin pump for at least six months, were in good general health with no conditions that could influence the outcome of the trial, and exercised at least 3  times per week performing 30 minutes of moderate or more vigorous aerobic activity. %Key exclusion criteria were pregnancy, one or more hypoglycemic episodes in the past 12 months requiring third party assistance for treatment, cardiovascular disease with inappropriate heart rate response to exercise or microvascular complications such as active proliferative retinopathy, use of agents that affect hepatic glucose production and use of pramlintide. 

\begin{figure}[h!]
        \centering
        \includegraphics[trim=3.8cm 0.5cm 3.8cm 1cm,clip,width=0.5\textwidth]{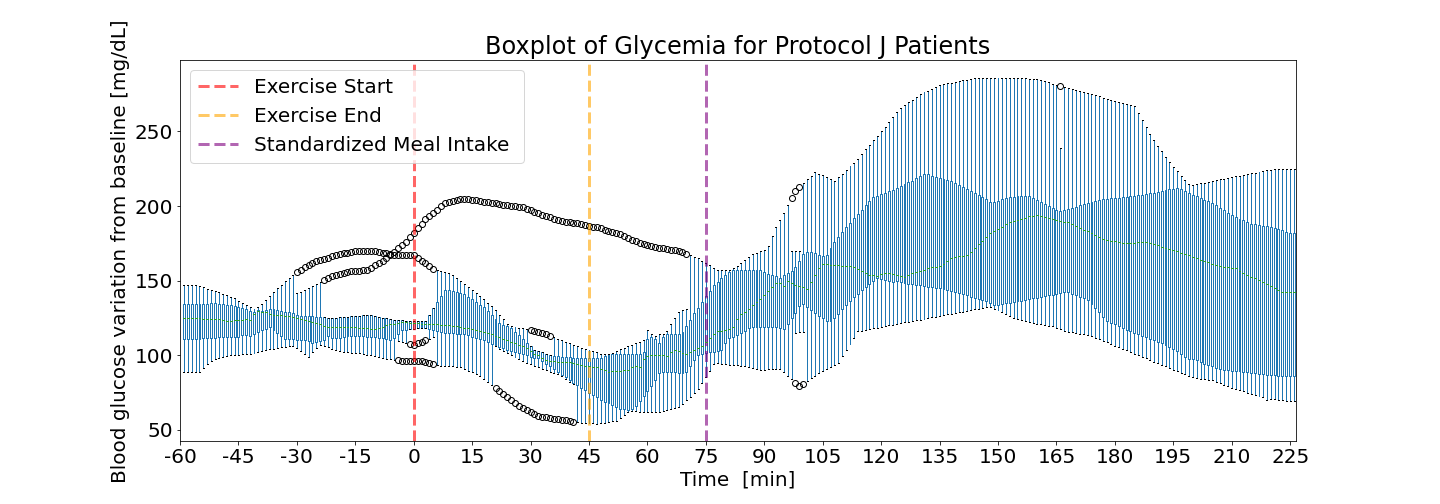}
        \caption{Actual glucose concentration levels [mg/dl] vs. Time [min]. Each boxplot represent data over the considered population, where the central mark in green is the median while the edges are 25th and 75th percentiles, respectively. \textit{Dashed red}: start of exercise session; \textit{Dashed yellow}: end of exercise session; \textit{Dashed purple}: meal intake.}
    \label{fig:control_arm}
\end{figure}

\begin{figure}
        \centering
        \includegraphics[trim=4.5cm 12cm 4.5cm 12cm,clip,width=0.5\textwidth]{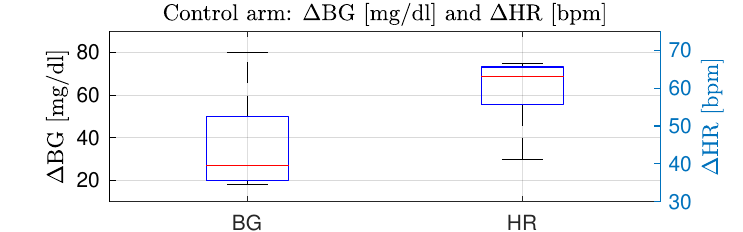}
        %\caption{Control arm: variation in glycemia [mg/dl] and heart rate [bpm] between start and end of the activity session, respectively. \textit{Top}: population boxplots, where the central mark is the median while the edges are 25th and 75th percentiles, respectively. \textit{Bottom}: variation in glycemia [mg/dl] plotted as function of heart rate [bpm] for all the patients in the population.}
        \caption{\textit{Control arm}: variation in glycemia [mg/dl] and heart rate [bpm] between start and end of the activity session, respectively. Population boxplots, where the central mark is the median while the edges are 25th and 75th percentiles, respectively.}
    \label{fig:BG_and_HR}
    \vspace{-0.5cm}
\end{figure}

After the screening visit, during which informed consent was obtained and individual VO$_2$max was assessed for the determination of the exercise intensity for the experimental trials, each participant underwent 4 aerobic exercise sessions (in random order) in the clinic, with different strategies for glucose regulation, as reported in Table \ref{tab:interventions}. 
\begin{table}%[b!]
  \caption{Interventions during the activity sessions}
\label{tab:interventions}
  \centering
  \begin{tabular}{ c| l }
  Arm $\#$ & Intervention\\
   \hline \hline
Control &  None\\
Strategy $1$&  Basal insulin reduction to 50$\%$\\
Strategy $2$& Dextrose tabs (20 [g]) orally 5 min prior and at 30 min\\
Strategy $3$& Glucagon (150 [$\mu$g]) subcutaneously 5 min prior \\
\hline 
\end{tabular}
\vspace{-0.5cm}
\end{table}
Each arm consisted in a 45-minutes moderate intensity aerobic activity (treadmill jogging/brisk walking) performed in the fasted state at $~50-55\%$ of the participants predetermined aerobic capacity. %The day before each session, a new insulin infusion set and a new continuous glucose monitoring (CGM) sensor were initiated. At the start of the exercise, participant blood glucose (BG) level must be 100 [mg/dl]$<$ BG $<$ 140 [mg/dl], and no injection of fast acting insulin must have occurred within 3 hours from the start. 
Continuous heart rate (HR [beats/min]) measurements were conducted along with CGMs. After the exercise session, participants rested for 30 minutes and then were given a standardized meal containing 44-50 [g] of carbohydrates. A meal bolus calculated on the basis of each individual insulin-to-carbohydrate ratio was administered 5 minutes prior to the meal intake. Paticipants were monitored for at least 2 hours after the meal and thereafter they were discharged from the clinic. %Each exercise session was separated by at least 3 days and were all completed by the participants within 12 weeks from the time of the screening visit. 
In this work, data from the \textit{Control arm} was used for our modeling and identification purposes, as described in Sec. \ref{sec:modeling}, while data from \textit{Strategy 1} was used for validation and verification purposes (see Sec. \ref{sec:results}). Fifteen subjects (9 M/6 F, median age of 30 years (IQR 25–43 years), BMI of 24 $kg/m^2$ (IQR 23–27 $kg/m^2$), T1D duration of 22 years (IQR 14–31 years), HbA1c of 6.8$\%$ (IQR 6.5-6.7$\%$)) completed the study, but records from 12 patients only were available to us at the time of writing. Due to incomplete data records or missing samples in the CGM during at least one of the exercise sessions, the records of 3 out of the 12 subjects were discarded. Further, in our analysis, we ratained the data collected from 1 hour before the in-hospital visit to 2 hours and 30 minutes after the meal intake. Figure~\ref{fig:control_arm} shows glucose traces for the considered population, while Fig.~\ref{fig:BG_and_HR} reports the maximum variation in glycemia [mg/dl] and heart rate [bpm] between start and end of the activity session, respectively.%, in the \textit{Control arm}. 
%
%\begin{figure*}[h!]
%        \centering
%       \includegraphics[trim=5cm 0cm 4cm 0cm,clip,width=\textwidth]{protocolJ_data_boxplot_control_trial_15min_xticks_offset_adjusted_2.png}
%        \caption{This is the caption.}
%    \label{fig:ds}
%\end{figure*}
%
%
%%%%%%%%%%%%%%%%%%%%%%%%%%%%%%%%%%%%%%%%%%%%%%%%%%%%%%%%%%%%%%%%%%%%%%%%%%%%%%%%
\section{Models of Glucose Dynamics Affected by Physical Activity} \label{sec:modeling}
\subsection{First-principles modeling}\label{sec:first_principles}
In order to accelerate the development of enhanced glucose control algorithms by performing \textit{in-silico} trials and to verify feasibility and effectiveness of novel treatment strategies on a population of virtual patients prior to their deployment
in clinical studies, we propose a mathematical model of the glucoregulatory system in people with diabetes capable of reproducing the glucose perturbations caused by moderate intensity aerobic activity. The model we propose is an extension to %a well-known simulation model of the
%glucose-insulin control system during a meal in T1D first proposed by Dalla Man and
%co-workers in
~\cite{dallaMan2007,DallaManChiara2007Gsso}. In brief, the model is comprised of a gastro-intestinal tract subsystem describing meal digestion, absorption of the carbohydrate content and subsequent glucose rate of appearance in blood; a glucose subsystem which describes insulin-independent glucose utilization by the brain and the erythrocytes which is constant and takes place in the first compartment (plasma and rapidly equilibrating tissues) and insulin-dependent glucose uptake $U_{id}$ [mg/kg/min] in the remote compartment (slowly equilibrating tissues); endogenous glucose production (EGP) in the liver; a subcutaneous insulin infusion module to simulate insulin transit from the subcutaneous space to plasma and finally an insulin subsystem representing the absorbed insulin in liver and plasma. More details about the complete metabolic model can be found in~\cite{dallaMan2007,DallaManChiara2007Gsso}, suffices it here to highlight that the primary effect of PA on glucose dynamics is to enhance insulin-dependent glucose utilization by
the tissues~\cite{carter, dallaMan2009}, which in nominal conditions is given by the following expression~\cite{dallaMan2007,DallaManChiara2007Gsso}:
\begin{equation}
U_{id}(t)= \frac{V_{m0} + V_{mX}\cdot X(t)}{k_{m0}+G(t)} \cdot G(t) \label{eq:uid}
\end{equation}
where $G_t(t)$ [mg/kg] is glucose mass in the perifery, $X(t)$ [pmol/l] is insulin action on glucose utilization:
\begin{equation}
\dot{X}(t) = -p_{2U} \cdot X(t) + p_{2U} [I(t)-I_b], \qquad X(0)= 0
\end{equation}
%%%
\begin{figure}[b!]
        \centering
        \vspace{-0.5cm}
        \includegraphics[trim=5cm 10cm 5cm 10cm,clip,width=0.4\textwidth]{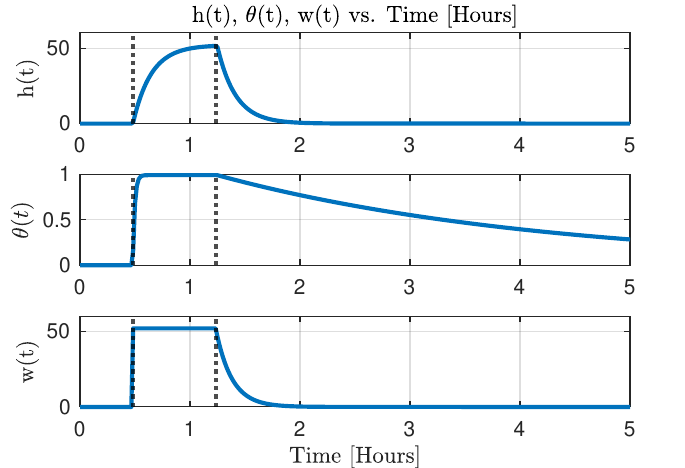}
        \caption{Evolution of $h(t)$, $\theta(t)$ and $w(t)$ vs Time [Hours] for $u_{hr}(t)$ equal to a step of magnitude 52 [bpm] and $\beta=0.052$ [bpm$^{-1}$], $\gamma=1.2$, $\varepsilon=0.01$, $\tau_h=10$ [min], $\kappa= 0.1151$ [min] and $\tau_{\theta}=180$ [min].}
    \label{fig:metabolic_variables}
\end{figure}
%%%
\noindent $I(t)$ represents plasma insulin concentration, the suffix $b$ denoting basal state, and $V_{m0}$ [mg/kg/min], $V_{mX}$ [mg/kg/min per pmol/l], $k_{m0}$ [mg/kg] and $p_{2U}$ [min$^{-1}$] are model parameters. In particular,  $V_{mX}$ denotes \textit{insulin sensitivity}, i.e., the ability of insulin to stimulate glucose utilization, and $p_{2U}$  denotes the rate constant of insulin action on peripheral glucose utilization. In this work, in order to account for the metabolic changes in glucose physiology during and after PA, we propose a set of variations to Eq.\eqref{eq:uid}, which are motivated by findings and observations reported in the medical literature. We make the assumption that the activity is of moderate intensity and aerobic, and therefore the main substrate metabolism for the required energy is that of glucose. Further, we assume that glucose metabolism during exercise depends on its duration and intensity, and that PA intensity correlates with heart rate, following~\cite{breton2008, dallaMan2009}.  In contrast to healthy subjects, BG has been observed to decline during prolonged exercise in people with diabetes \cite{martin1995,minuk1981} caused by impaired EGP and concomitant increase in plasma glucose uptake by the skeletal muscles \cite{borghouts}. %During exercise, glucose transport into the muscle cells is enhanced and facilitated by the GLUT4 transporters \cite{james32,goodyear33}. 
Immediately after the start of exercise, the number of glucose transporter-4 (GLUT4) proteins on the plasma membrane and transverse tubules dramatically increases, associated with muscle contractions~\cite{goodyear41,goodyear43,roy104}. This leads to an increased glucose transport capacity into the muscle cells, and an increase in the rate of glucose uptake~\cite{james32,goodyear33}. Within 2 hours after exercise is discontinued, however, GLUT4 levels return to resting values~\cite{goodyear41}. We have modeled these rapid mechanisms and short-term effects causing a drop in BG concentration in people with diabetes, by raising insulin-independent glucose clearance after the onset of exercise as a function of exercise intensity measured by heart rate: %multiplying $V_{m0}$ by $1+\beta h(t)$, where $\beta$ is an individual-specific parameter and $h(t)$ is a function of the net difference in heart rate from baseline, which slowly decays after the end of the exercise and becomes negligible after 2 hours: 
$V_{m0}(1+\beta h(t))$, where $\beta$ is an individual-specific parameter and $h(t)$ is a delayed function of the net difference in heart rate from baseline, paralleling~\cite{breton2008}, which slowly decays after the end of the exercise and becomes negligible after 2 hours. In addition, we have enhanced glucose utilization by lowering insulin action: $k_{m0}(1-\varepsilon w(t))$, where $\varepsilon$ is a constant and $w(t)$ is a function of the heart rate, which remains steady during the session and exponentially decays right after to become negligible after 2 hours. In the post-exercise period, more GLUT4 translocates to the plasma membrane upon insulin stimulation, possibly involving mechanisms initiated by glycogen depletion~\cite{borghouts}, causing prolonged effects on glucose uptake through an increased insulin sensivity for at least 16 hours post exercise bout. To capture this, we have introduced the term $V_{mX}(1+\gamma \theta(t))$, $\gamma$ is individual-specific and $1<\gamma<2$ in agreement with published literature reporting insulin sensitivity to almost double due to exercise \cite{schiavon}, and $\theta(t)$ is a function reaching the value of 1 very rapidly at the beginning of exercise and slowly decreasing over 16 hours to replicate the findings in~\cite{borghouts}. Based on these observations, we propose the following equation to describe $U_{id}$ accounting for PA:
\begin{equation}\label{eq:uid_exercise}
U_{id}(t)= \frac{V_{m0}(1+\beta h(t))+ V_{mX}(1+\gamma \theta(t)) \cdot X(t)}{k_{m0}(1-\varepsilon w(t))+G(t)} \cdot G(t)
\end{equation}
where:
\begin{align}
\dot{h}(t) &= -\frac{1}{\tau_{h}}(h(t)-u_{hr}(t)), \: u_{hr}(t)= \text{HR}(t)-\text{HR}_{b} \label{eq:upa}
\\
\dot{\theta}(t)&= -(\varphi(t)+\frac{1}{\tau_{\theta}}) \theta(t) + \varphi(t) \label{eq:theta}
\\
\varphi(t) &= \frac{u_{hr}(t)}{1+u_{hr}(t)} \label{eq:varphi}
\\
w(t) &=
  \begin{cases} \label{eq:kappa}
    0        & \quad \text{prior to exercise} \\
    u_{hr}(t)  & \quad \text{during exercise} \\
    u_{hr}(t_{\text{end}}) e^{-\kappa t} & \quad \text{after exercise} 
  \end{cases} 
  \end{align}
and $u_{hr}(t)$ is an external input measuring exercise intensity and duration based on HR [bpm], HR$_{b}$ indicating baseline before the start of exercise, initial conditions are $h(0)=0$, $\theta(0)=0$, $\varphi(0)=0$, and finally, $\beta$ [bpm$^{-1}$], $\gamma$, $\varepsilon$, $\tau_{h}$ [min], $\kappa$ [min] and $\tau_{\theta}$ [min] are model parameters. Fig.~\ref{fig:metabolic_variables} illustrates the evolution over time of the functions $h(t)$, $\theta(t)$ and $w(t)$ for a nominal step-like heart rate increase $u_{hr}(t)=52$ [bpm] and $\beta=0.052$ [bpm$^{-1}$], $\gamma=1.2$, $\varepsilon=0.01$, $\tau_h=10$ [min], $\kappa= 0.1151$ [min] and $\tau_{\theta}=180$ [min].
%We have tuned the time constants $\tau_{h}$, $\kappa$, $\tau_{\theta}$ according to the physiological considerations presented above, while $\varepsilon$ and $\gamma$ were chosen empirically. Table~\ref{tab:params} reports the paramenter values considered herein, while Fig.~\ref{fig:metabolic_variables} illustrates the evolution of $h(t)$, $\theta(t)$ and $w(t)$ over time for $u_{hr}(t)$ equal to a step of magnitude 52 [bpm]. Last, we elaborate on the procedure followed for the estimation of $\beta$ in Sec.~\ref{sec:estimation}.
%\begin{table}%[h!]
%  \caption{Physical Activity Model Parameters}
%\label{tab:params}
%  \centering
%  \begin{tabular}{c| c |c }
%  Parameter & Value & Unit \\
%   \hline \hline
%$\gamma$    & 1.2 &  Dimensionless \\
%$\varepsilon$    & 0.01 &  Dimensionless  \\
%$\tau_{h}$    & 10 & [min]  \\
%$\kappa$    & 0.1151 &  [min]  \\
%$\tau_{\theta}$    & 180 &  [min] \\             
%\hline 
% \end{tabular}
%\end{table}
%
%
%\begin{figure}[h!]
%        \centering
%        %\includegraphics[trim=7cm 0 7cm 0,clip,width=0.5\textwidth]{protocolJ_data_boxplot_control_rial_15min_xticks.png}
%        \includegraphics[trim=3.5cm 9cm 3.5cm 9cm,clip,width=0.5\textwidth]{metabolic_variables_new_5hours_adjusted.pdf}
%        \caption{Evolution of $h(t)$, $\theta(t)$ and $w(t)$ vs Time [Hours] for $u_{hr}(t)$ equal to a step of magnitude 52 [bpm] and $\beta=0.052$ [bpm$^{-1}$], $\gamma=1.2$, $\varepsilon=0.01$, $\tau_h=10$ [min], $\kappa= 0.1151$ [min] and $\tau_{\theta}=180$ [min].}
%    \label{fig:metabolic_variables}
%\end{figure}
%------------------------------------
\subsection{Control-oriented modeling}
The metabolic model equations describing PA outlined in Sec.~\ref{sec:first_principles} constitute a very useful tool to reproduce real-life scenarios in simulation. However, they are not appealing to be used for control design due to their complexity. Hence, we propose a simpler lower order model able to capture the main dynamics of glycemia during an exercise bout from the perspective of controller synthesis. Specifically, we propose a model able to predict the glucose change during PA in response to a step increase in heart rate, to be used as a model of the PA disturbance. While in real life heart rate may change with several patterns, our assumption of a step increase mimicks the case of a constant moderate effort aerobic activity, and thus it is valid within the scope of our paper. That said, denoting with $u_{hr}(t)$ [bpm] the input and $y_{BG}(t)$ [mg/dl] the output, the following relation is derived:
\begin{equation}\label{eq:BG_drop}
\dot{y}_{BG}(t) = - \frac{1}{T} [y_{BG}(t)+ K \cdot u_{hr}(t-\tau)]
\end{equation}
where $T$ [min] is the time constant, $\tau$ [min] is the time delay between input and output, and $K$ [mg/dl/bpm] is the gain. Taking Laplace transform of Eq.~\eqref{eq:BG_drop}, one obtains:
\begin{equation}
Y_{BG}(s) = G(s) \cdot U_{hr}(s)
\end{equation}
where $G(s)$ is given by
\begin{equation}\label{eq:TF}
G(s) = -e^{-s\tau} \frac{K}{T\cdot s+1}
\end{equation}
The parameters $\tau$, $K$ and $T$ are individual specific and can be estimated from actual data, in a grey-box identification fashion.
\subsection{Parameter estimation}\label{sec:estimation}
We have tuned the time constants $\tau_{h}$, $\kappa$, $\tau_{\theta}$ in Eqs.~\eqref{eq:upa}--\eqref{eq:kappa} according to the physiological considerations and findings presented in Sec.~\ref{sec:first_principles}, while $\varepsilon$ and $\gamma$ were chosen empirically. While we acknowledge the implicit limitations in the use of a nominal parameter set, the lack of suitable experimental data supporting system identification, i.e., data collected with the triple-tracer technique, allowing the estimation of the glucose fluxes between compartments, at the time of writing, prevented a data-driven individual specific parameter estimation procedure. We would like to remind the reader, however, that the development of a validated parsimonious simulator of glucose metabolism during PA, equipped with validated virtual subjects, was extraneous to the scope of this work. Our objective in this contribution was to propose a description of the relationship between PA and changes in crucial parameters such as insulin sensitivity, glucose utilization and ultimately BG dynamics capable of mimicking as closely as possible that of the actual diabetes population, allowing us to replicate \textit{in-silico} previous protocols and simulate novel treatment strategies and insulin dosing algorithms for the AP on a variety of real-life scenarios. 
%The identification of the unknown parameters in Eqs.~\eqref{eq:uid_exercise}, \eqref{eq:theta}, \eqref{eq:varphi} and \eqref{eq:kappa} requires experimental data collected with the triple-tracer technique, allowing the estimation of the glucose fluxes between compartments. At the time of writing, the lack of this type of data precludes the possibility of validating the proposed model. While we
%acknowledge the limitations, we would like to emphasize that the development of a validated digital twin of a patient during physical activity was not the scope of this paper, since our objective was to propose models able to capture as closely as possible the dynamics behavior of blood glucose during exercise at a population level, to enable \textit{in-silico} testing of novel insuling dosing algorithms for the artificial pancreas.

That said, we have introduced some personalization in our model via the parameter $\beta$ in Eq.~\eqref{eq:uid_exercise}. For its estimation we used a prediction error method (PEM) with
quadratic cost~\cite{ljung1999}:
\begin{equation}\label{eq:PEM}
\hat{\xi} = \text{arg} \min_{\xi} \frac{1}{N} \sum_{i=1}^{N}(y(i)-\hat{y}(i,\xi))^2 
\end{equation}
where $\xi=\beta$, $y(i)$ denotes actual BG level collected during the \textit{Control arm} of the trial, for each of the available patients, from the start of the exercise session up until the end, $\hat{y}(i,\xi)$ is the simulated BG level obtained with the metabolic model, reproducing the same \textit{in-silico} protocol of the \textit{Control arm}, and $N$ is the number of available data points. For the exercise model we used the nominal parameter set discussed hereinabove, while for the gastro-intestinal tract subsystem, glucose subsystem, EGP, insulin subsystem and subcutaneous insulin kinetics, we used the parameter values from the 10 \textit{in-silico} adults in \cite{kovatchevBorisP2008MSAC}. Input $u_{hr}(t)$ was taken from actual available data. Initial conditions were $\hat{\beta}_0=0.005$ [bpm$^{-1}$]and $\hat{y}_0=125$ [mg/dl].

As for the unknown coefficients in the transfer function model given in Eq.~\eqref{eq:TF}, their estimation was carried out as follows. First, the time delay $\tau$ was derived by visual inspection of the BG time series from the \textit{Control arm}. Next, with the newly estimated time delay, and as input a step of amplitude $\max(\text{HR}(t)-\text{HR}_b)$ and duration that of the exercise session, the coefficients $\xi = [K \quad T]^\intercal$ were calculated fitting model~\eqref{eq:TF} to the simulated $\hat{y}(\xi)$ adopting a PEM with cost given in Eq.~\eqref{eq:PEM}. 
%%%%%%%%%%%%%%%%%%%%%%%%%%%%%%%%%%%%%%%%%%%%%%%%%%%%%%%%%%%%%%%%%%%%%%%%%%%%%%%%
%\section{A decision support system for automatic titration and insulin dosing}\label{sec:control_design}
%\subsection{Current standard of care}
%\subsection{Control objectives and controller design}
%%%%%%%%%%%%%%%%%%%%%%%%%%%%%%%%%%%%%%%%%%%%%%%%%%%%%%%%%%%%%%%%%%%%%%%%%%%%%%%%
\section{Results} \label{sec:results}
\subsection{\textit{In-silico} experiments}
The nominal parameter set for our proposed exercise model in Eqs.~\eqref{eq:upa}--\eqref{eq:kappa}, is reported in Table~\ref{tab:params}, while the individual specific $\hat{\beta}$ is given in Table~\ref{tab:beta}.
%\subsection{Parameter estimation}
%\begin{table}[b!]
%  \caption{Physical Activity Model Parameters}
%\label{tab:params}
%  \centering
%  \begin{tabular}{c| c |c }
%  Parameter & Value & Unit \\
%   \hline \hline
%$\gamma$    & 1.2 &  Dimensionless \\
%$\varepsilon$    & 0.01 &  Dimensionless  \\
%$\tau_{h}$    & 10 & [min]  \\
%$\kappa$    & 0.1151 &  [min]  \\
%$\tau_{\theta}$    & 180 &  [min] \\             
%\hline 
% \end{tabular}
%\end{table}
%%
%\begin{table}[b!]
%  \caption{Parameter $\beta$ [bpm$^{-1}$]}
%\label{tab:beta}
%  \centering
%  \begin{tabular}{c| c}
%  Patient ID$\#$ & Value \\
%   \hline \hline
%  1  & 0.0143 \\
%  2  & 0.0566 \\
%  3  & 0.1016 \\
%  4  & 0.0173 \\
%  5  & 0.0144 \\     
%  6  & 0.0149 \\
%  7  & 0.1655 \\
%  8  & 0.0521 \\
%  9  & 0.0446 \\
%  median(25th,75th) &  0.0046(0.0148,0.0678)\\             
%\hline 
% \end{tabular}
%\end{table}
%%\subsection{\textit{In-silico} experiments}
\begin{figure}[b!]
\centering
\includegraphics[trim=3.8cm 0cm 3.8cm 1cm,clip,width=0.5\textwidth]{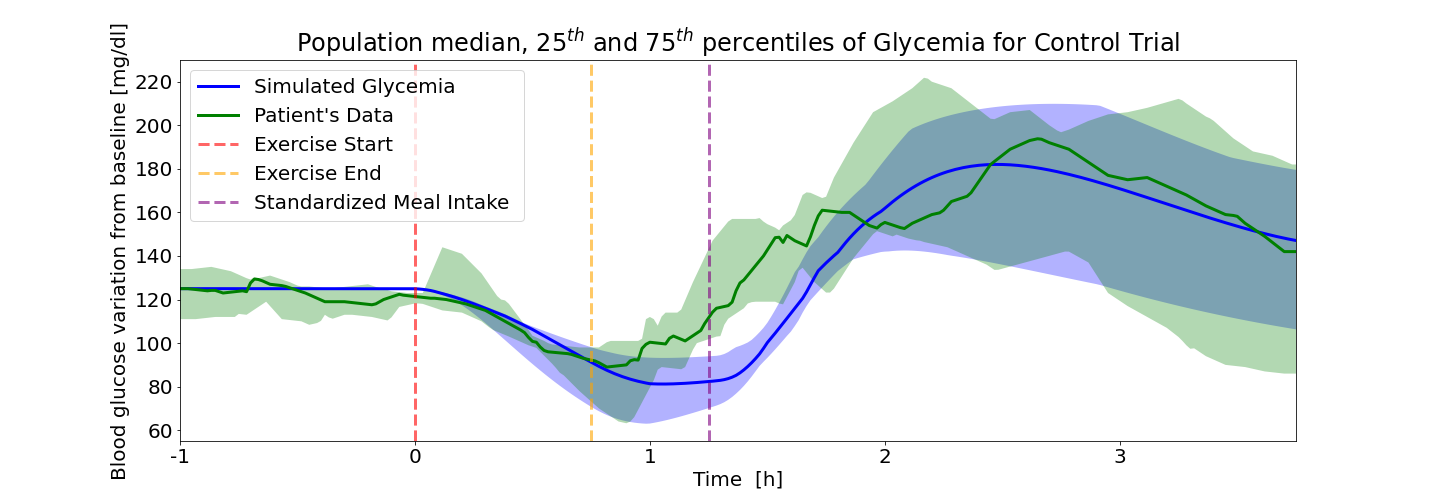}
\includegraphics[trim=3.8cm 0.5cm 3.8cm 0cm,clip,width=0.5\textwidth]{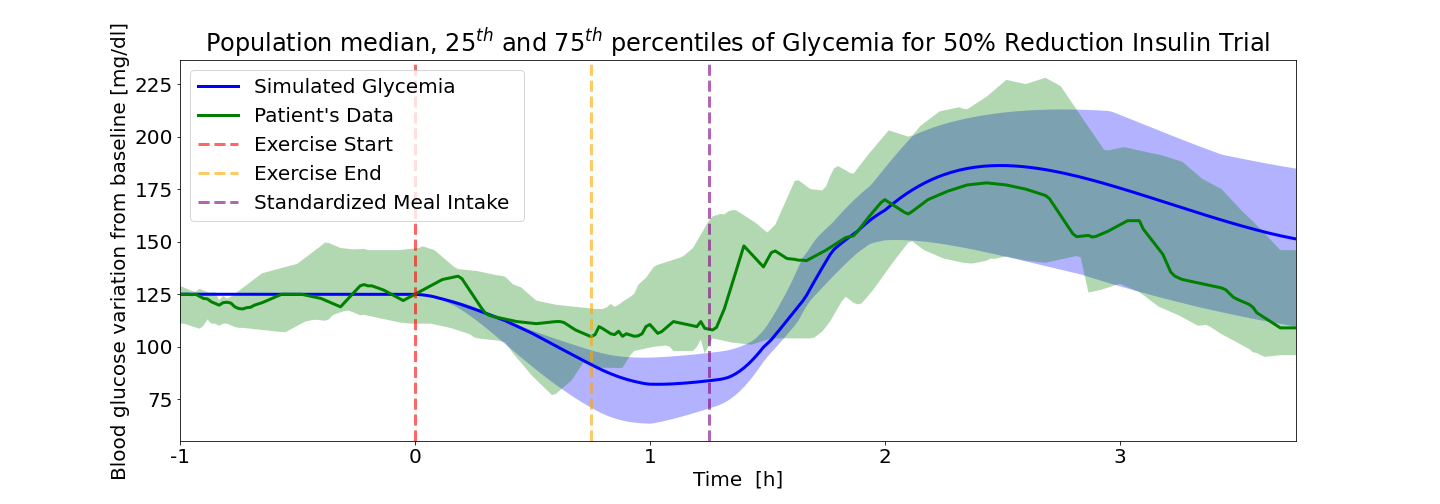}
%\includegraphics[trim=3.8cm 0cm 3.8cm 0cm,clip,width=0.5\textwidth]{compariosn_tabs_trial_percentiles_simulated_data_tabs_24.png}
%compariosn_50_reduction_trial_percentiles_color_simulated_50reduc_data4
%compariosn_control_trial_percentiles_simulated_data_44
%compariosn_tabs_trial_percentiles_simulated_data_tabs_24
%
        \caption{Comparison between actual patient data (\textit{Green}) and simulated data (\textit{Blue}) vs. Time [hours]. The thick line is the population median, the shaded area represents the range between 25th and 75th percentiles. \textit{Dashed red}: start of exercise session; \textit{Dashed yellow}: end of exercise session; \textit{Dashed purple}: meal intake. \textit{Top}: Control, \textit{Bottom} Strategy 1.}%, \textit{Bottom} Strategy 2.}
    \label{fig:experiments}
\end{figure}
To verify the correctness of our proposed extension to the metabolic model, we performed \textit{in-silico} simulations to reproduce, at a population level, the glucose traces collected during the \textit{Control} and \textit{Strategy 1} arms of the previously described clinical trial, succintely reported in Table~\ref{tab:interventions}. We generated a population of 10 virtual subjects with T1D, where parameter values from the 10-\textit{in-silico} adults in \cite{kovatchevBorisP2008MSAC} were used for the gastro-intestinal tract subsystem, glucose subsystem, EGP, insulin subsystem and subcutaneous insulin kinetics, while values in Tables \ref{tab:params} and \ref{tab:beta} were used for the parameter set in Eqs.~\eqref{eq:uid_exercise}--\eqref{eq:kappa}. Last, the input $u_{hr}$ to Eq.~\eqref{eq:upa}, was assumed to be a step function with amplitude matching that of the actual patient data. Figure~\ref{fig:experiments} shows the actual patients data vs. the simulated ones. In each panel, the thick lines denote the median, while the shaded areas are the range between 25th and 75th percentiles, respectively. Before the start of the sessions, identified by the red dashed line in the panels, the large inter-patient variability in the actual population is apparent for each of the study arm, as expected. This was not the case for the simulated data because all our simulations started with the same initial fasting BG value of 125 [mg/dl]. From the top panel, comparing results over the \textit{Control arm}, we can see that during the session, the simulated data are well in agreement with the actual data, meaning that our model is able to reproduce the impact of moderate intensity aerobic activity on BG dynamics. Moreover, our model is able to accurately capture the nadir of the BG time-series. In the actual population, glycemia seem to return almost to its pre-activity value within 30 minutes from the end of the session, while still fasting. Our model exhibits a time lag in the recovery post-exercise compared to the true data. This time delay may be attributed to a mismatch of the EGP model for the window of time immediately after activity is suspended.  We remind the reader in passing the without \textit{ad-hoc} experiments enabling collection of glucose tracers concentration, it is not possible to estimate glucose fluxes between compartments, and hence it is not possible at this stage to make any assumptions or draw conclusions on the change in EGP dynamics. The simulated data follow the actual data with good accuracy after the meal intake, confirming the long-term change in insulin sensitivity post-exercise that we have introduced. Similar observations can be made by inspection of the bottom panel, which presents results pertaining to \textit{Strategy 1}. Simulated data follows the actual data well during the activity and after the meal, however shows a delayed response immediately after exercise. 
%, reported in Table \ref{tab:upa}.
\begin{table}%[b!]
  \caption{Physical Activity Model Parameters}
\label{tab:params}
  \centering
  \begin{tabular}{c| c |c }
  Parameter & Value & Unit \\
   \hline \hline
$\gamma$    & 1.2 &  Dimensionless \\
$\varepsilon$    & 0.01 &  Dimensionless  \\
$\tau_{h}$    & 10 & [min]  \\
$\kappa$    & 0.1151 &  [min]  \\
$\tau_{\theta}$    & 180 &  [min] \\             
\hline 
 \end{tabular}
\end{table}
\begin{table}%b!]
  \caption{Parameter $\beta$ [bpm$^{-1}$]}
\label{tab:beta}
  \centering
  \begin{tabular}{c| c}
  Patient ID$\#$ & Value \\
   \hline \hline
  1  & 0.0143 \\
  2  & 0.0566 \\
  3  & 0.1016 \\
  4  & 0.0173 \\
  5  & 0.0144 \\     
  6  & 0.0149 \\
  7  & 0.1655 \\
  8  & 0.0521 \\
  9  & 0.0446 \\
  median(25th,75th) &  0.0046(0.0148,0.0678)\\             
\hline 
 \end{tabular}
 \vspace{-0.5cm}
\end{table}
\begin{table}[h!]
  \caption{Transfer function parameter estimates}
\label{tab:TF_estimates}
  \centering
  \begin{tabular}{c|c|c|c}
  Patient ID$\#$ & $\tau$ [min]& $K$ [mg/dl/bpm]& $T$ [min]\\
   \hline \hline
  1  & 10 & 0.7334 & 51.3\\
  2  & 15 & 2.714 & 72.3\\
  3  & 15 & 3.946 & 103.8\\
  4  &  10& 5.272 & 549.5\\
  5  & 10 & 0.7402 & 62.8\\     
  6  & 15 & 1.6950 & 22.7 \\
  7  & 15 & 5.1460 & 535.0\\
  8  & 10 & 1.8150 & 105.4\\
  9  & 15 & 5.5840 & 652.0\\
  median(IQR) &  15(10,15) & 2.72(1.45,5.17) & 103.80(59.92,538.62) \\             
\hline 
 \end{tabular}
\end{table}
\subsection{Data-driven transfer function identification}
\begin{figure}
        \centering
        \includegraphics[trim=5cm 9.5cm 5cm 10.5cm,clip,width=0.5\textwidth]{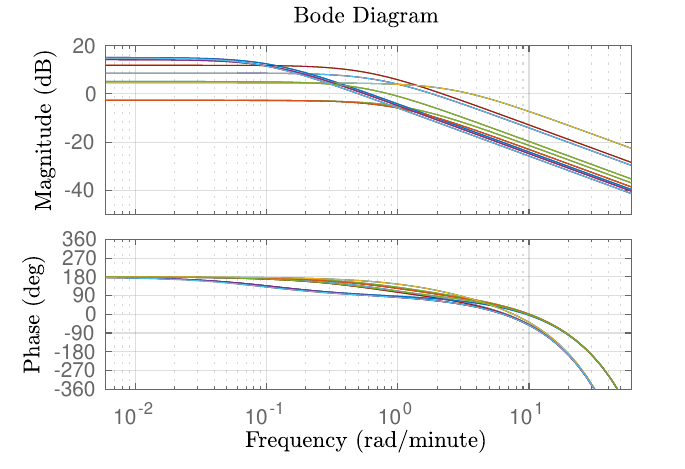}
        \caption{Bode diagrams of identified transfer function models.}
    \label{fig:bode}
\end{figure}

\begin{figure}
        \centering
        \includegraphics[trim=5.8cm 12cm 6cm 12cm,clip,width=0.4\textwidth]{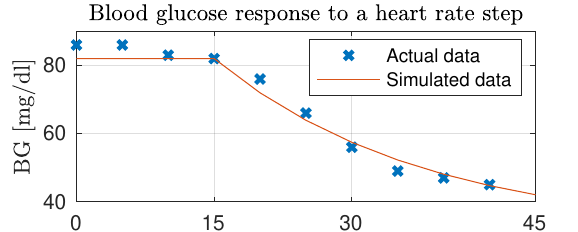}
        \includegraphics[trim=5.8cm 12cm 6cm 11.5cm,clip,width=0.4\textwidth]{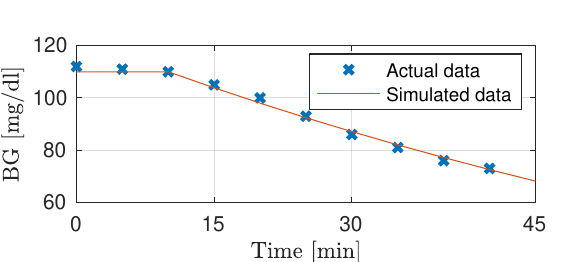}
        \caption{\textit{Control arm}, blood glucose level during the activity session. \textit{Blue star} Actual data [mg/dl], \textit{Orange Line} Simulated data [mg/dl] vs. Time [min]. The simulated data was obtained with the transfer function in Eq.~\eqref{eq:TF} and parameters in Table~\ref{tab:TF_estimates}. \textit{Top} Patient $\#$6, \textit{Bottom} Patient $\#$8.}
    \label{fig:patient18}
\end{figure}

Table~\ref{tab:TF_estimates} presents the values of the unknown coefficients in the transfer function model of Eq.~\eqref{eq:TF} identified from patients data, while Fig.~\ref{fig:bode} depicts the Bode diagrams of each transfer function. A comparison between actual and simulated BG concentration, for two representative patients, is given in Fig.~\ref{fig:patient18}. We investigated the accuracy of the identified model in terms of FIT and Root Mean Square Error (RMSE), defined as follows:
\begin{equation}
e_{\text{FIT}} = \text{max} \Big( 0,1-\frac{\lVert \hat{y}_{BG}-y_{BG} \rVert}{\lVert y_{BG} - \bar{y}_{BG} \rVert_2} \Big)
\end{equation}
\begin{equation}
e_{\text{RMS}} = \sqrt{\frac{1}{N} \sum_{i=1}^N \lVert y_{BG}(i)-\hat{y}_{BG}(i)\rVert^2} 
\end{equation}
where $\hat{y}_{BG}$ is the simulated BG profile during the exercise session obtained feeding our proposed transfer function with a step in heart rate as input, $y_{BG}$ are the true BG values, $\bar{y}_{BG}$ is the mean of $y_{BG}$ and $N$ is the number of samples. The results are reported in Table~\ref{tab:performances} and show very good predictive capabilities of our simple model. 
%\begin{figure}
%        \centering
%        \includegraphics[trim=5cm 9cm 5cm 10cm,clip,width=0.5\textwidth]{patient18ident.pdf}
%        \caption{Caption}
%    \label{fig:patient18}
%\end{figure}
%
%
\begin{table}[t!]
  \caption{Accuracy of transfer function model predictions}
\label{tab:performances}
  \centering
  \begin{tabular}{c|c|c}
  Patient ID$\#$ & $e_\text{FIT}$ & $e_\text{RMS}$ [mg/dl]\\
   \hline \hline
  1  & 0.8861 &  0.7243\\
  2  & 0.9370 &  8.1852\\
  3  & 0.9499 &  1.2090\\
  4  & 0.8541 &  1.5913\\
  5  & 0.8155 &  1.8981\\     
  6  & 0.8775 &  2.5980\\
  7  & 0.7980 &  10.27\\
  8  & 0.9152 &  1.2018\\
  9  & 0.7184 &  3.1144\\
  median(IQR) &  0.8775(0.8111,0.9206) & 1.8981(1.2072,4.3821)\\             
\hline 
 \end{tabular}
 \vspace{-0.5cm}
\end{table}
%
%\section{Implications for the artificial pancreas}\label{sec:control_design}
%\subsection{Current standard of care}
%\subsection{Control objectives and controller design}
%
\section{Summary and Conclusions}\label{sec:conclusions}
One of the major obstacles to the development of a fully automated AP to be used by the patients in daily life conditions, is the adaptation of insulin therapy during PA. The availability of models able to predict the effect of exercise on glucose disposal accurately and reliably constitutes the first step toward realizing such a fully automated closed-loop system. In this work, we proposed a physiology-based parsimonious model of activity, building on the work of~\cite{breton2008,dallaMan2009}. We incorporated our proposed model into a metabolic simulator of the glucose-insulin system~\cite{dallaMan2007} to obtain a simulation tool able to reproduce actual data collected from people with T1D during a clinical trial. To the best of our knowledge, this is the first effort of this kind, where the prediction capabilities of the candidate physiological model with respect to glucose dynamics both during and after exercise and meal intake are directly compared to actual patients data. While parsimonious models represents a very useful tool to simulate real-life scenarios, their value is limited in the context of control design due to their complexity. Motivated by this, we proposed simple transfer function models able to replicate BG response to a step in HR. The use of the HR signal as an external input makes the proposed models suitable for a practical implementation, since HR can be easily measured \textit{in-vivo} with wearable devices. Validation of the models on an independent dataset was not possible at this stage, and it is left to future work. Moving forward, we plan on integrating the transfer functions in a decision support tool to be used as hypoglycemia predictor during an exercise bout and to modulate insulin therapy accordingly, by means of a model-based controller. We intend to use our proposed simulator to verify our closed-loop strategy \textit{in-silico}.
%\addtolength{\textheight}{-12cm}   % This command serves to balance the column lengths
                                  % on the last page of the document manually. It shortens
                                  % the textheight of the last page by a suitable amount.
                                  % This command does not take effect until the next page
                                  % so it should come on the page before the last. Make
                                  % sure that you do not shorten the textheight too much.

%%%%%%%%%%%%%%%%%%%%%%%%%%%%%%%%%%%%%%%%%%%%%%%%%%%%%%%%%%%%%%%%%%%%%%%%%%%%%%%%
%\section*{APPENDIX}
%
%Appendixes should appear before the acknowledgment.

%\section*{ACKNOWLEDGMENT}

%This work was supported by the University of Houston through a startup grant.% <-this % stops a space
%%%%%%%%%%%%%%%%%%%%%%%%%%%%%%%%%%%%%%%%%%%%%%%%%%%%%%%%%%%%%%%%%%%%%%%%%%%%%%
%------------------------------------------
\bibliographystyle{IEEEtran} % use IEEEtran.bst style
\bibliography{CCTA23_final}
%=================================

\end{document}